\documentclass[12pt,preprint]{aastex}
\usepackage{amsmath,longtable,natbib}
\usepackage{lineno}
\usepackage{float}
\usepackage[caption = false]{subfig}
\usepackage{hyperref}
\usepackage{graphicx}
\shorttitle{GHRSS $-$ update on the survey}
\shortauthors{Bhattacharyya et al.}
\usepackage{subfig}

\begin{document}

\title{The GMRT High Resolution Southern Sky Survey for pulsars and transients -II. New discoveries, timing and polarization properties}
\author{
B.~Bhattacharyya\altaffilmark{1},
J.~Roy\altaffilmark{1},
B.~W.~Stappers\altaffilmark{2},
T.~Johnson\altaffilmark{3},
C.~D.~Ilie\altaffilmark{2},
A.~Lyne\altaffilmark{2},
M.~Malenta\altaffilmark{2},
P.~Weltevrede\altaffilmark{2},
J.~Chengalur\altaffilmark{1},
S.~Cooper\altaffilmark{1},
B.~Kaur\altaffilmark{1},
M.~Keith\altaffilmark{1},
M.~Kerr\altaffilmark{4},
S.~Kudale\altaffilmark{1},
M.~A.~McLaughlin\altaffilmark{5,6},
S.~M.~Ransom\altaffilmark{7},
P.~S.~Ray\altaffilmark{4}}
\altaffiltext{1}{National Centre for Radio Astrophysics, Tata Institute of Fundamental Research, Pune 411 007, India}
\altaffiltext{2}{Jodrell Bank Centre for Astrophysics, School of Physics and Astronomy, The University of Manchester, Manchester M13 9PL, UK}
\altaffiltext{3}{College of Science, George Mason University, Fairfax, VA 22030, resident at Naval Research Laboratory, Washington, DC 20375, USA}
\altaffiltext{4}{Space Science Division, Naval Research Laboratory, Washington, DC 20375-5352, USA}
\altaffiltext{5}{Department of Physics \& Astronomy, West Virginia University, Morgantown, WV 26506, US}
\altaffiltext{6}{Center for Gravitational Waves and Cosmology, West Virginia University, Chestnut Ridge Research Building, Morgantown, WV 26505}
\altaffiltext{7}{National Radio Astronomy Observatory(NRAO), Charlottesville, VA 22903, USA}

\affil{}

\begin{abstract}
We have been conducting the GMRT High Resolution Southern Sky (GHRSS) survey for the last four years and have discovered 18 pulsars to date. 
The GHRSS survey is an off-Galactic-plane survey at 322 MHz in a region of the sky (declination range $-$40\degr~ to $-$54\degr) complementary to other 
ongoing low-frequency surveys. In this paper we report the discovery of three pulsars, PSRs J1239$-$48, J1516$-$43 and J1726$-$52. 
We also present timing solutions for three pulsars previously discovered with the GHRSS survey: PSR~J2144$-$5237, a millisecond pulsar 
with a period $P=5$~ms in a 10 day~orbit around a $\leq$ 0.18 M$_\sun$~companion; PSR J1516$-$43, a mildly recycled $P=36$~ms pulsar in a 228 day~orbit with a companion of mass $\sim$0.4 M$_\sun$; and the $P=320$~ms PSR~J0514$-$4408 which we show is a source of pulsed $\gamma$-ray emission.  
We also report radio polarimetric observations of three of the GHRSS discoveries, PSRs J0418$-$4154, J0514$-$4408 and J2144$-$5237. 

\end{abstract}

\vskip 0.6 cm

\section{Introduction}
\label{sec:intro}

Neutron stars are accessible to observations as pulsars and provide a valuable means for probing the behaviour of matter, 
energy, space and time under extraordinarily diverse conditions. Studies of normal pulsars 
having spin period $>$30~ms can reveal interesting properties like glitches, profile state changes, 
nulling and intermittency (e.g. \cite{lyne96}, \cite{kramer06}). The extreme stability of the spin of 
millisecond pulsars (MSPs) makes them ideal laboratories to test the physics of gravity \citep{lee12}. 
In spite of the fact that the rates of discovery of pulsars in ongoing surveys at major telescopes over the last decade have 
increased dramatically, the presently known population of about 2600 pulsars is less than 5\% of the predicted number of detectable 
radio pulsars \citep{faucher06}. A large fraction of the pulsars are faint sources requiring sensitive searches 
and improved analysis techniques for discovery.  
Pulsar surveys are sensitivity limited, hence the design of more sensitive instruments promises a higher discovery rate. Large 
arrays of many smaller telescopes is one possible strategy for sensitivity improvement and is implemented in the Giant Metrewave Radio Telescope 
(GMRT\footnote{http://gmrt.ncra.tifr.res.in}). It is the largest array telescope at metre wavelengths and has the potential to 
undertake sensitive pulsar searches, a potential which was confirmed with the discovery of 23 pulsars in targeted and blind searches (\cite{bh13},
\cite{bh16}). We have been carrying out the GMRT High Resolution Southern Sky (GHRSS) survey using the 32 MHz bandwidth 
GMRT Software Backend (GSB, \cite{roy10}) for pulsars and transients since the fall of 2013. In this paper, the 32 MHz bandwidth 
component of the GHRSS survey will be notated as GHRSS ``phase-1''. The GHRSS phase-1 is an off-Galactic-plane ($|b|$~$>$~5\degr) 
survey at 322 MHz whose declination range of $-$40\degr~ to $-$54\degr~ complements other ongoing 
low-frequency surveys with the GBT\footnote{http://arcc.phys.utb.edu/gbncc/} and LOFAR\footnote{http://www.astron.nl/lotaas}. The survey 
description and initial discovery of 10 pulsars are reported in \citet[][hereafter P1]{bh16}.
Beginning in late 2017, we embarked on phase-2 of this survey with the upgraded GMRT using up to 200 MHz of bandwidth 
(Roy et al. 2018). The survey description and discoveries with the GHRSS phase-2 will be reported in a follow up paper. 

Following the discovery of a pulsar, the next essential step is regular timing to characterize its nature, 
rotation properties, and companion type and orbital properties if in a binary system. 
Precise localistion of the newly discovered pulsars by the GMRT interferometric array reduces the discovery positional 
uncertainty of $\pm$ 40\arcmin~to the size of the synthesized beam of the array, $\pm$ 10\arcsec~(\cite{roy13}, \cite{roy12}). 
This allowed us to carry out more sensitive follow up observations with the narrower coherent array beam.  

Radiation from pulsars is believed to originate from the particles streaming outward along the open field lines above the poles of an 
essentially dipolar magnetic field. Linear polarization at any point in the profile is related to the orientation of the magnetic field 
at the corresponding point of origin. In the simplest form the position angle (PA) of the linear polarization within the pulse window rotates smoothly 
as a function of longitude in an ``S'' shaped fashion described within the ``rotating vector model'' \citep{rc69}. 
Studying the polarization of radio pulsars is important for understanding the geometry and underlying emission mechanisms. We performed polarization
studies of three of the pulsars discovered with GHRSS phase-1 survey with the Parkes telescope.
\begin{table*}
\begin{center}
\caption{Parameters of the pulsars discovered in the GHRSS survey and studied in this paper}
\vspace{0.3cm}
\label{discovery}
\begin{tabular}{|l|c|c|c|c|c|c|c|c|c|c|c|c|c|c|c|c}
\hline
Pulsar name    & Period  & Dispersion measure & Detection significance & Flux density$^\dagger$ \\
               & (ms)    & (pc~cm$^{-3}$)            & ($\sigma$)             & (mJy)          \\\hline
PSR J0418$-$4154 & 757.11  & 24.325(9)             & 50                     & 10.3                       \\
PSR J0514$-$4408 & 320.27$^\ddagger$  & 15.122(6)             & 42                     & 9.7                     \\
{\bf PSR J1239$-$48} & 653.9    & 107.6            & 21                  & 0.4                  \\
{\bf PSR J1516$-$43} & 36.02   & 70.3              & 9                      & 0.7                \\
{\bf PSR J1726$-$52} & 631.8  & 119.7              & 8                      & 0.7                \\
PSR J2144$-$5237 & 5.04    & 19.5465(2)               & 9                    & 1.6                    \\\hline
\end{tabular}
\end{center}
 We announce the discovery of the pulsars marked in bold face in this paper.\\
 Uncertainty in dispersion measure value in the last digit are quoted in the parentheses for the pulsars for which accurate measurement is possible with long-term timing\\
$^\dagger$ : Flux density is without primary beam correction for the three newly discovered pulsars.\\
$^\ddagger$ : Please note that there was a typo in the period mentioned for this pulsar in P1.\\
\vspace{1cm}
\end{table*}

Section \ref{sec:obs} of this paper details the search and timing observations with the GMRT and polarization observations with Parkes. We 
describe the discovery of three pulsars, PSRs J1239$-$48, J1516$-$43 and J1726$-$52, in Section \ref{sec:discoveries}. 
Section \ref{sec:timing} details the timing study of PSR J0514$-$4408 (originally reported as J0514$-$4407 in P1), mildly recycled 
pulsar J1516$-$43, and MSP J2144$-$5237. Section \ref{sec:lat_0514} details the detection of $\gamma$-ray pulses from 
PSR~J0514$-$4408 with the \textit{Fermi} Large Area Telescope (LAT). Section \ref{sec:polarization} describes the polarization properties of three of 
the GHRSS pulsars, J0418$-$4154, J0514$-$4408 MSP J2144$-$5237. In Section \ref{sec:discussion} we present 
discussion of the results and the summary. 

  \begin{figure}[!ht]
    \subfloat[\label{fig:512ch}]{%
      \includegraphics[width=2in,angle=-90]{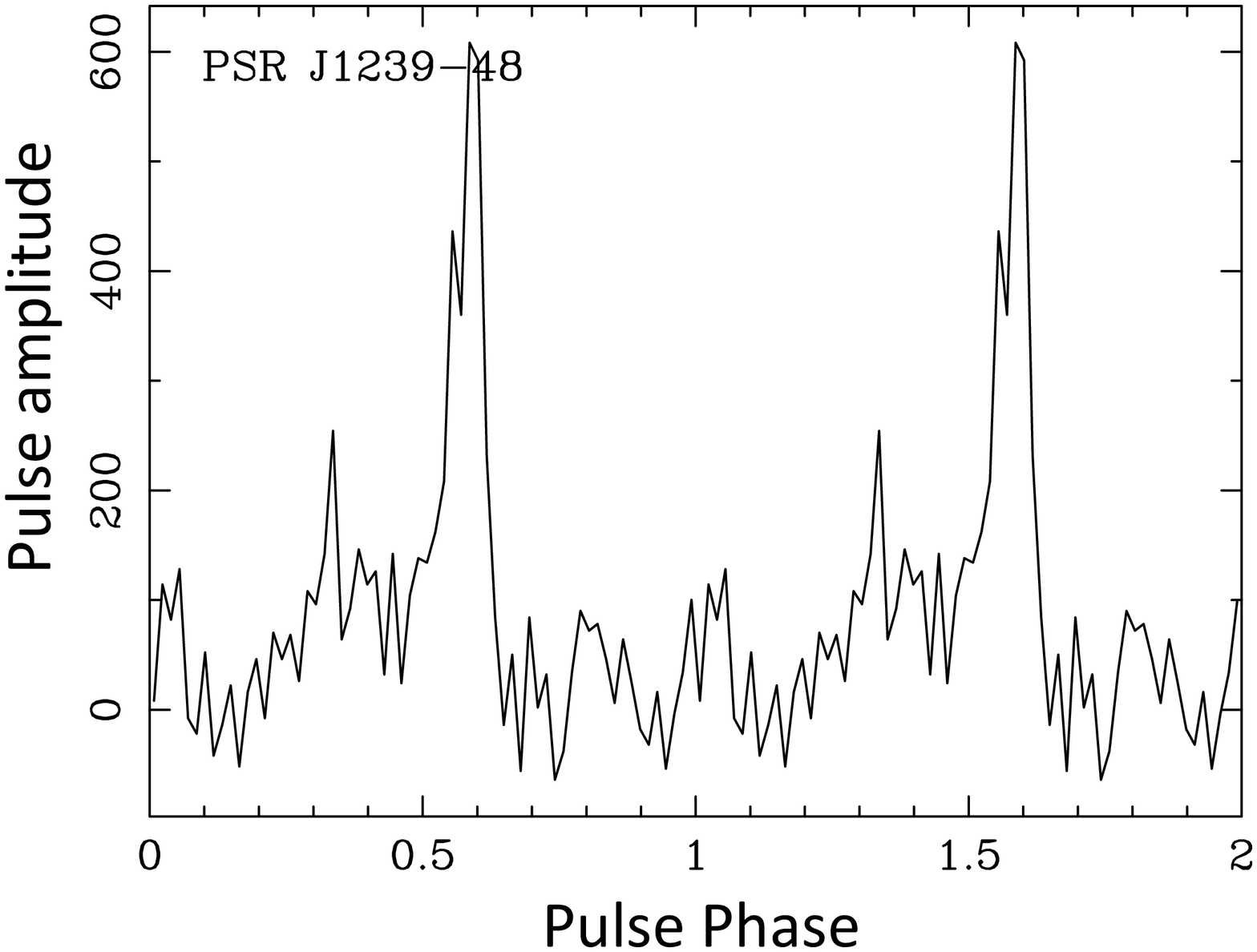}
    }
    \hfill
    \subfloat[\label{fig:2048ch}]{%
      \includegraphics[width=2in,angle=-90]{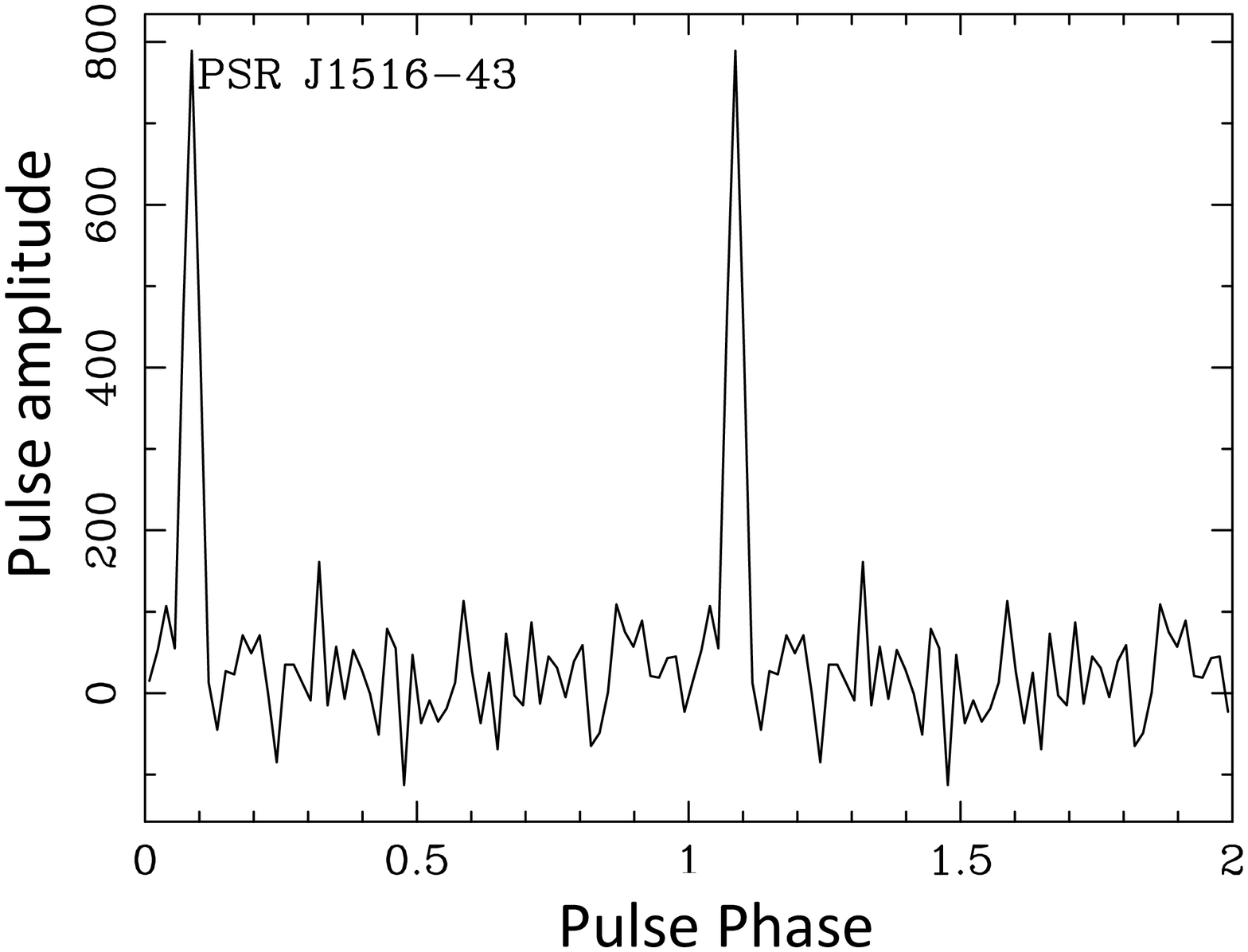}
    }
    \subfloat[\label{fig:2048ch}]{%
      \includegraphics[width=2in,angle=-90]{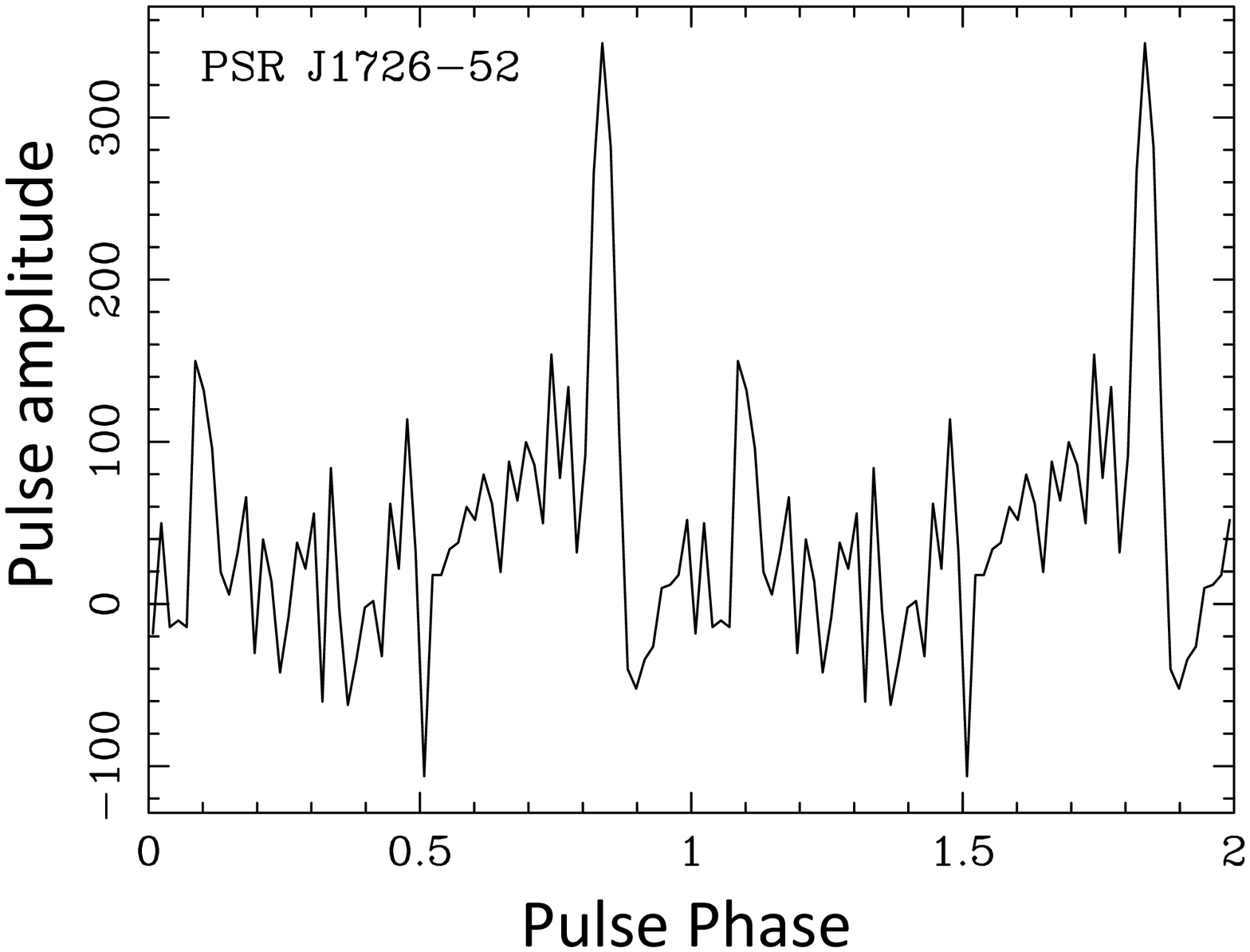}
    }
    \caption{Pulse profiles (two spin periods) of three newly discovered pulsars with the GHRSS survey with 15 minutes integration at a centre frequency of 322 MHz and bandwidth of 32 MHz: (a) PSR J1239$-$48, (b) J1516$-$43, (c) J1726$-$52. Pulse amplitude is in arbitrary units.}
    \label{fig_discovery}
  \end{figure}

\begin{figure}
\centering
\begin{tabular}{c c}
\subfloat{\includegraphics[scale=0.34,angle=-90]{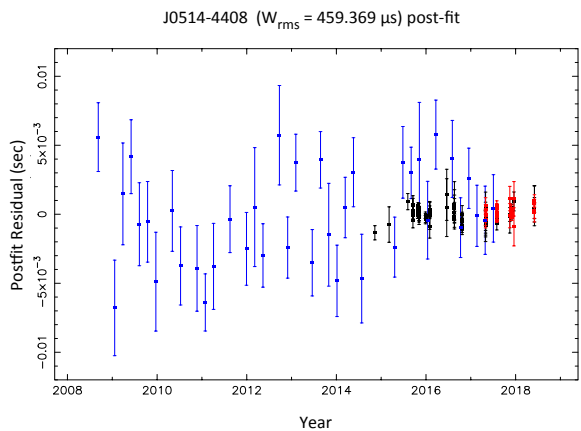}}
\\
\multicolumn{2}{c}
\subfloat{\includegraphics[scale=0.34,angle=-90]{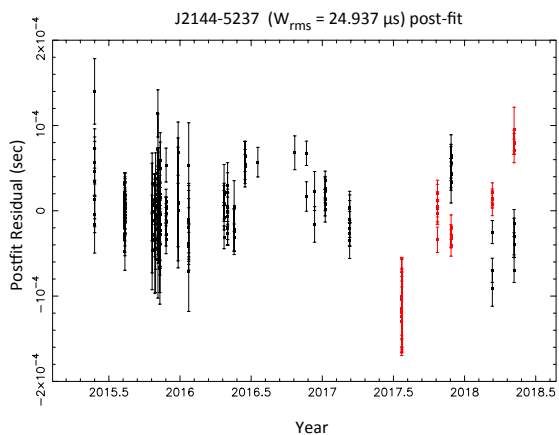}}
\end{tabular}\caption{Top panel: Combined radio and $\gamma$-ray timing residuals for PSR J0514$-$4408. The black points represent residuals at 322 MHz with bandwidth of 32 MHz using the GMRT legacy system, the red points represent residuals at 400 MHz with a bandwidth of 200 MHz using the upgraded GMRT, the blue points represent the $\gamma$-ray timing residuals from \textit{Fermi} LAT. Bottom panel: Radio timing residuals for PSR J2144$-$5237 from the GMRT observations, black and red points as in top panel.}
\label{fig:timing_residuals}
\end{figure}

\section{Observation and analysis}
\label{sec:obs}
\begin{table*}
\centering
\caption{Timing parameters of PSR J0514$-$4408 and J2144$-$5237
\label{tab:residuals}}
\begin{tabular}{|l|c||c|c|c|}
\hline
Name &J0514$-$4408 &J2144$-$5237\\
  \hline
  \multicolumn{3}{c}{Gated imaging position$^\ast$} \\
  \hline
 Right ascension (J2000)\dotfill &05$^\mathrm{h}$14$^\mathrm{m}$51$^\mathrm{s}$.84(1$^\mathrm{s}$.04) &21$^\mathrm{h}$44$^\mathrm{m}$39$^\mathrm{s}$.2(65$^\mathrm{s}$.7) \\
 Declination (J2000)\dotfill     &$-$44\degr07\arcmin06\farcs51(8\farcs4)         &$-$52\degr37\arcmin32\farcs17(3\farcs8) \\
  \hline
  \multicolumn{3}{c}{Parameters from radio and $\gamma$-ray timing$^\ast$} \\
  \hline
 Right ascension (J2000)\dotfill &05$^\mathrm{h}$14$^\mathrm{m}$52\fs190(3)        &21$^\mathrm{h}$44$^\mathrm{m}$35\fs65(6) \\
 Declination (J2000)\dotfill     &$-$44\degr08\arcmin37\farcs38(2)                 &$-$52\degr37\arcmin07\farcs53(2)\\
 Pulsar frequency $f$ (Hz)\dotfill   &3.122357486324(6)                           &198.3554831467(9)\\
 Pulsar frequency derivative $\dot{f}$ (Hz s$^{-1}$)\dotfill &$-$1.99080(1)$\times$10$^{-14}$  &$-$3.50(2)$\times$10$^{-16}$\\
 Period epoch (MJD)\dotfill           &57330                                      &57328\\
 Dispersion measure $\mbox{DM}$$^\dagger$ (pc~cm$^{-3}$)\dotfill &15.122(6)                   &19.5465(2)\\
 Binary model\dotfill & $-$                   & ELL1\\
 Orbital period $P_{b}$ (days)\dotfill&$-$&10.5803185(2) \\
 Projected semi-major axis $x$ (lt-s)\dotfill &$-$&6.361098(1) \\
 Epoch of ascending node passage $T_{\rm {ASC}}$ (MJD)\dotfill &$-$               &57497.785577172346066(1)\\
 Timing Data Span                                              &54715.2$-$58271.5 & 57167.9$-$58245.1\\
 Number of TOAs\dotfill  &155               &217\\
 Reduced Chi-square \dotfill       & 1.4                       & 2.9  \\
 Post-fit residual rms (ms)\dotfill       & 0.459                       & 0.024  \\\hline
 \multicolumn{3}{c}{Derived parameters} \\
  \hline
 Period (ms) \dotfill                                          & 320.270822408985(6)         & 5.04145377851813(2)                   \\
 Period Derivative (s/s) \dotfill                                   & 2.04203(2)$\times$10$^{-15}$ & 8.89(7)$\times$10$^{-21}$                  \\
 Total time span (yr) \dotfill                                & 9.7                          & 2.9                  \\
 Spin down energy loss rate $\dot{E}$ (erg/s)\dotfill         & 2.4$\times$10$^{33}$         & 2.7$\times$10$^{33}$       \\
 Characteristic age (yr)\dotfill                              & 2.5$\times$10$^{6}$          & 8.9$\times$10$^{9}$ \\
 Surface magnetic flux density (Gauss)\dotfill                & 8.2$\times$10$^{11}$         & 2.1$\times$10$^{8}$ \\
 Rotation measure (rad m$^{-2}$)\dotfill                      & 17.3                         & 25.1 \\
 DM distance (kpc)$^\ddagger$ \dotfill                        & 0.8                          & 0.8\\
 DM distance (kpc)$^\ddagger$$^\dagger$ \dotfill              & 0.9                          & 1.6\\\hline
\end{tabular}
\\
$^\ast$ Errors correspond to 1$\sigma$.\\
$^\dagger$ DM values are calculated from fitting sub-band TOAs from 300$-$500 MHz wide observing band of uGMRT\\
$^\ddagger$ using the \cite{cl01} model of electron distribution\\
$^\ddagger$$^\dagger$ using the \cite{yao17} model of electron distribution\\ 
We note that the calculated DM distance is model dependent.\\
Timing uses DE421 solar system ephemeris.\\
\end{table*}

The observing setup for GHRSS phase-1 is detailed in P1. The GMRT is a multi-element aperture synthesis telescope 
consisting of 30 antennas, each 45~m diameter, spread over a 25 km-diameter region and operating at 5 frequencies ranging 
from 150 MHz to 1450 MHz \citep{swarup97}. The observations used the GMRT Software Backend, a fully real-time backend utilizing
an FX correlator\footnote{http://www.gmrt.ncra.tifr.res.in/gmrt\_hpage/Users/doc/WEBLF/LFRA/node76.html} and a beamformer for an array 
of 32 dual polarized signals Nyquist sampled at 33 or 66 MHz \citep{roy10}.
For the survey observations we used $\sim$ 61.44 $\mu$s time resolution with $\sim$ 16.275 kHz frequency resolution over 32 MHz 
observing bandwidth for mid-Galactic latitudes and $\sim$ 30.72 $\mu$s time resolution with $\sim$ 32.55 kHz frequency resolution over 32 MHz
observing bandwidth for high-Galactic latitudes. A factor of two better frequency resolution is used at mid-Galactic latitudes to compensate 
for larger dispersion smearing. The calculated theoretical search sensitivity                                              
for a 15 min GHRSS pointing with incoherent array gain of 2.5 K/Jy at 322 MHz is 0.5 mJy for a 5$\sigma$ detection assuming a  
10\% duty cycle and a total system temperature at 322 MHz of 106 K (P1; for minimum T$_{sky}$ of GHRSS). 
As demonstrated in P1 (Fig. 10), the observed GHRSS survey sensitivity is within $\pm$ 50\% of the theoretical one, 
thus allowing us to detect faint pulsars. 
We recorded Stokes-I at a data rate of 32 MB/s for 8 bit samples. We used the wider incoherent beam 
of the GMRT (FWHM of 80\arcmin~at 322 MHz), which is ideal for blind pulsar surveys. 
With 60\% of the GHRSS survey ($\sim$1800 deg$^2$) we have collected about 30 TB of data. We searched for pulsations 
using a 512 core cluster (10 Tflops) at the National Centre for Radio Astrophysics (NCRA) and a {\sc presto}$-$based \citep{Ransom02} pipeline. 
The dedispersion range used in the search is 0 to 500 pc cm$^{-3}$ (discussed in P1).
We used an acceleration search allowing for up to 5 m s$^{-2}$ line-of-sight acceleration for a 2 ms pulsar 
over 15 mins of observing duration and up to 8 harmonics were used in harmonic summing. Further details 
about the search analysis pipeline can be found in P1.
\begin{table*}
\centering
\caption{Timing parameters of PSR J1516$-$43
\label{tab:residuals_1516}}
\begin{tabular}{|l|c||c|c|c}
  \hline
Right ascension (J2000)\dotfill &15$^\mathrm{h}$16$^\mathrm{m}$32$^\mathrm{s}$31(1$^\mathrm{s}$.09)        \\
Declination (J2000)\dotfill     &$-$43\degr20\arcmin00\farcs00(1\farcs0)                 \\
Pulsar frequency $f$ (Hz)\dotfill   &27.760652(2)                           \\
Period epoch (MJD)\dotfill           & 57575.4                                    \\
Dispersion measure $\mbox{DM}$ (pc~cm$^{-3}$)\dotfill &70.3                   \\
Binary model$^\dagger$\dotfill  & BT \\
Orbital period $P_{b}$ (days)\dotfill& 228.4(1)\\
Projected semi-major axis $x$ (lt-s)\dotfill & 1.0(1)\\
Epoch of ascending node passage $T_{\rm {ASC}}$ (MJD)\dotfill &57575.419(1)\\
Number of TOAs\dotfill            & 40 \\
Reduced Chi-square \dotfill       & 1.04   \\\hline
\multicolumn{2}{c}{Derived parameters} \\\hline
Period (ms) \dotfill                 & 36.022209(2)\\
DM distance (kpc)$^\ddagger$ \dotfill & 1.8\\
DM distance (kpc)$^\ddagger\dagger$ \dotfill & 3.0 \\
Total time span (yr) \dotfill        & 520 days \\\hline
\end{tabular}
\\
$^\dagger$ used by {\sc fitorbit} software\\
$^\ddagger$ using the \cite{cl01} model of electron distribution\\
$^\ddagger$$^\dagger$ using the \cite{yao17} model of electron distribution\\
We note that the calculated DM distance is model dependent.\\ 
\end{table*}

\begin{figure}
\centering
\begin{tabular}{c c}
\subfloat{\includegraphics[scale=0.34,angle=-90]{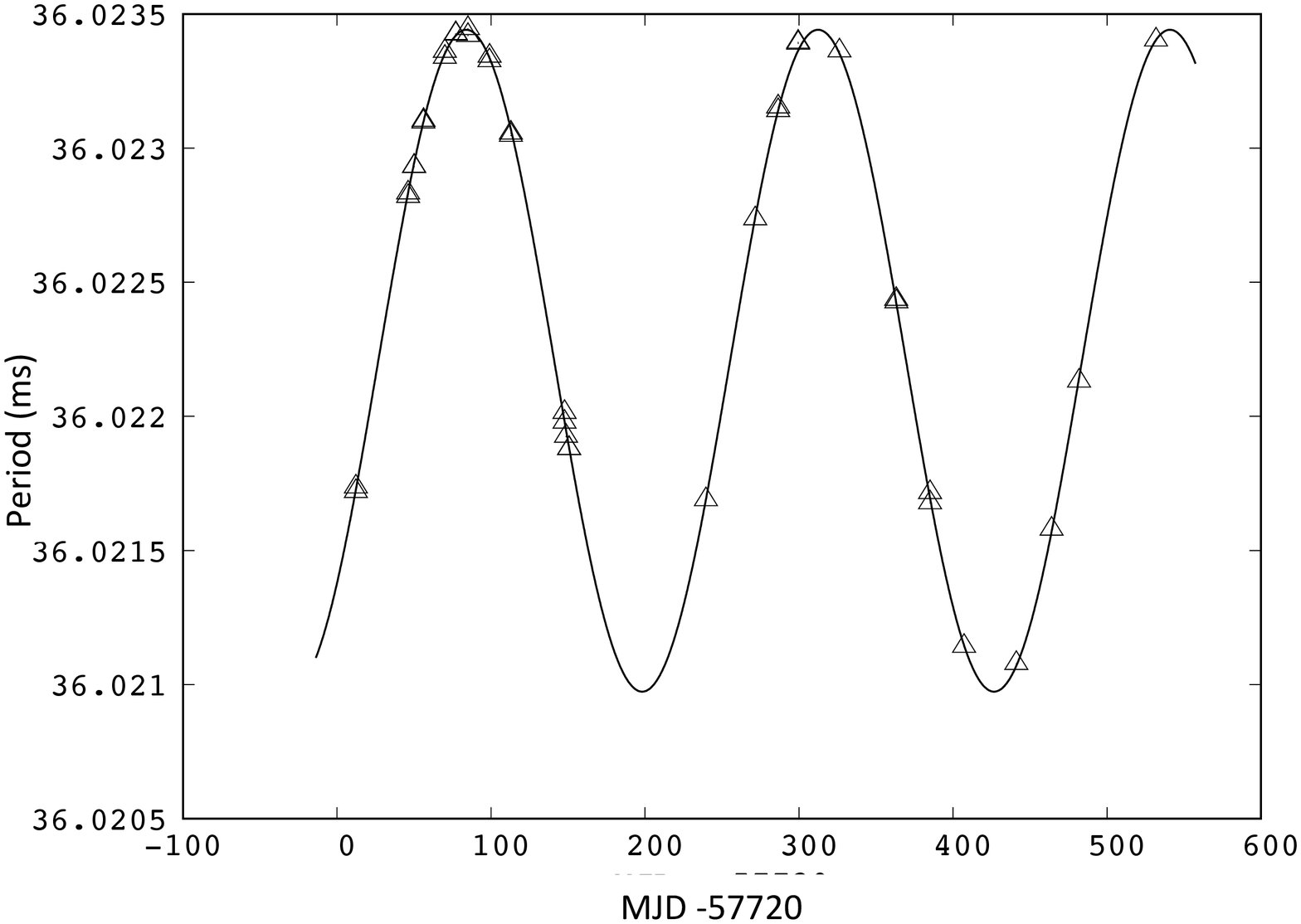}}
\\
\multicolumn{2}{c}
\subfloat{\includegraphics[scale=0.34,angle=-90]{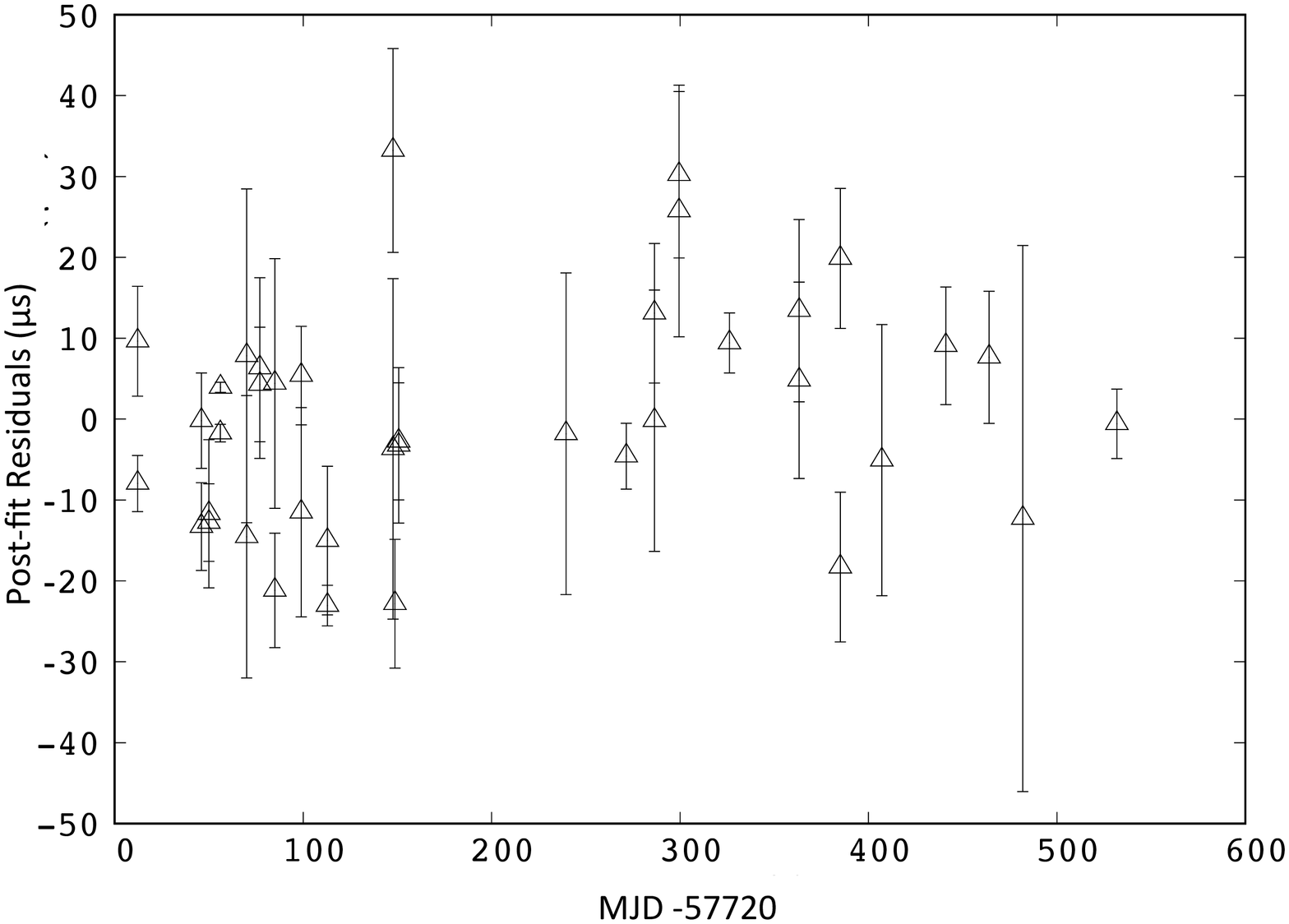}}
\end{tabular}\caption{Timing studies of the mildly recycled pulsar J1516$-$43. Top panel: Measured pulse period variation (triangles) and fitted orbital model (solid line), Bottom panel: Post-fit residuals.}
\label{fig:1516}
\end{figure}
We can localise the newly discovered pulsars and transients in the image plane with the GMRT interferometric array with an 
accuracy of better than $\pm$ 10\arcsec (half of the typical synthesized beam used in the image made at 322 MHz) using gated 
imaging of pulsars \citep{roy13}. For the pulsars which are localised in the image plane, we use the smaller field of view but more sensitive 
coherent array for follow up observations. Using the coherent array with the central core of the GMRT having 17 antennas 
(i.e. gain of $\sim$ 7 K/Jy) the timing sensitivity is 0.3 mJy for 10$\sigma$ detection. After discovery we started a 
regular timing campaign at 322 MHz. 
We used the highest signal-to-noise ratio profiles as templates for extracting times-of-arrival (TOAs). 
The TOAs are modeled using the standard pulsar timing software {\sc tempo2}\footnote{http://www.atnf.csiro.au/research/pulsar/tempo2}.

The polarimetric observations of the three GHRSS pulsars (J0418$-$4154, J0514$-$4408 and J2144$-$5237) were performed on the 10th of 
September 2017 with the Parkes radio telescope. We used the central beam of the multibeam receiver \citep{Staveley96} 
and the PDFB4 backend at a central frequency of 1369 MHz and a bandwidth of 256 MHz.  More 
details about the receiver and the backend used in this observation can be found in \cite{manchester13}. The receiver consists 
of two linear, perpendicular dipoles which receive the orthogonal components of the incoming electric field. These two fields were 
correlated to produce the four Stokes parameters. The first half of each observation was performed with a feed angle 
rotation of $-45^{\circ}$, and the second half with a feed angle rotation of $+45^{\circ}$, allowing asymmetries in the 
performance of the two signal paths corresponding to the two polarizations to effectively cancel out. In order to calibrate 
for the leakage between the dipoles, a polarimetric calibration model \citep{vs04} has been applied to the data, as derived 
for the Parkes Pulsar Timing Array project\footnote{https://www.atnf.csiro.au/research/pulsar/ppta/}. 
For some additional details about the methodology we refer to \cite{wj08}.

The observations were structured as follows: PSR J0418$-$4154 was observed for 3600 s, PSR J0514$-$4408 for 13500 s and  
PSR J2144$-$5237 for 12250 s (in folded mode with 30 s sub-integrations). A 120 s calibration 
observation with the noise diode switched on was performed before the first half and after the second half. Each half was calibrated 
using its corresponding polarization calibration observation. The two halves were added together to form the final integrated 
polarized profiles. The tools which were used to produce the plots in this section are part of the \textsc{PSRSALSA}\footnote{https://github.com/weltevrede/psrsalsa} software package \citep{wel16}, publicly available at the link provided.  
\begin{figure}
\begin{center}
\includegraphics[width=4in,angle=0]{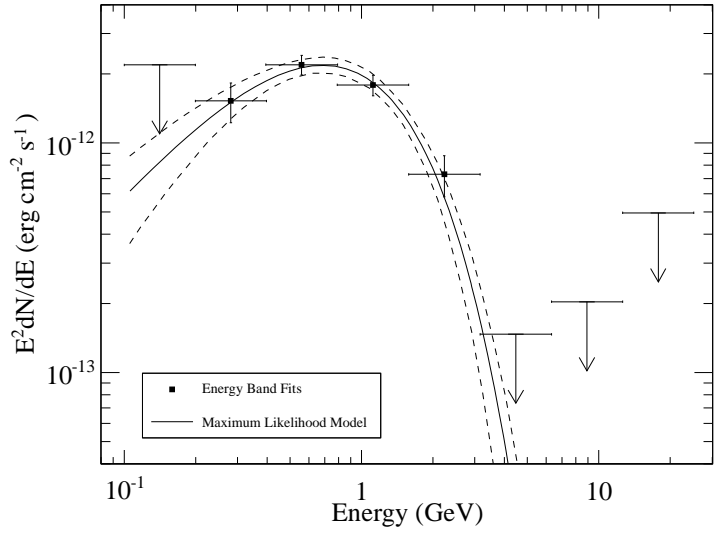}
\caption{Phase-averaged $\gamma$-ray spectrum of PSR~J0514$-$4408.  The solid line is the maximum-likelihood model and the dashed lines represent the $\pm1\sigma$ uncertainty from the fit. The points were derived from likelihood fits in the individual energy bands with the pulsar spectrum modeled as a power law with fixed photon index of 2. All uncertainties are statistical only. A 95\% confidence-level upper limit is plotted for those energy bands in which the pulsar was detected with TS$<$9 ($\sim3\sigma$) or with $<$4 predicted counts.}
\label{fig:latspectrum}
\end{center}
\end{figure}

\section{Results}
\label{sec:res}
\subsection{New discoveries}
\label{sec:discoveries}

In this paper we announce the discovery of PSRs J1239$-$48, J1516$-$43 and J1726$-$52 in GHRSS phase-1. 
Fig. \ref{fig_discovery} shows the discovery profiles of these three pulsars. Table \ref{discovery} presents the spin period, 
dispersion measure (DM) and flux density of these three pulsars marked in boldface. PSR J1239$-$48 is a 653.9 ms pulsar having 
a DM of 107.6 pc cm$^{-3}$ and estimated discovery flux density 0.4 mJy. PSR J1516$-$43 has a
period of 36.02 ms, a DM of 70.3 pc cm$^{-3}$, and a discovery flux density of 0.7 mJy. 
PSR J1726$-$52 is a 631.8 ms pulsar at a DM of 119.7 pc cm$^{-3}$ with a flux density 0.7 mJy. Detection of such faint pulsars 
with the GHRSS survey indicates that we are achieving our theoretical sensitivity limit of $\sim$ 0.5 mJy.

\subsection{Timing study}
\label{sec:timing}

We have been performing timing observations at approximately monthly cadence (with $\sim$ 15 minutes integration) since the discovery 
and derived timing solutions for PSRs J0514$-$4408 and J2144$-$5237. The timing solutions for these two pulsars are presented 
in Table \ref{tab:residuals}. We note that the timing position is off from the position derived from gated imaging. This 
could be due to the fact that these pulsars were observed at very low elevation angles (as the pulsars are very southern), 
so refractive effects can cause a significant shift between the measured position and the true position of the sources. Similar 
effects were also observed for other GHRSS discoveries reported in P1. The detection significance increase 
while pointing centre is at timing position.
PSR J0514$-$4408 is a 320 ms pulsar with a \textit{Fermi} LAT source 3FGL J0514.6$-$4406 $\sim$1.8\arcmin~ 
from the pulsar. Folding LAT photons with the radio timing model, we discovered $\gamma-$ray pulsations from this pulsar,
described in Section \ref{sec:lat_0514}. Fig. \ref{fig:timing_residuals} presents the radio and $\gamma-$ray timing 
residuals for PSR J0514$-$4408. 
We have generated 38 TOAs from the \textit{Fermi} LAT observations spanning 9.2 years, which are presented as blue 
points in Fig. \ref{fig:timing_residuals}. We checked the preliminary 8-year \textit{Fermi} LAT catalog\footnote{Available at \url{https://fermi.gsfc.nasa.gov/ssc/data/access/lat/8yr\_catalog/}} \citep{4FGL} for positional associations with the other pulsars in Table \ref{discovery} 
and found none.
PSR J2144$-$5237 is a 5.04 ms pulsar in a binary with orbital period of 10.6 days for which the timing model is presented in Table \ref{tab:residuals}
and timing residuals are plotted in Fig. \ref{fig:timing_residuals}. The calculated mass function \citep{lorimer04} of PSR J2144$-$5237 
is 0.002 M$_\sun$, which corresponds to a companion mass range of 0.18--0.46 M$_\sun$ considering 90\degr~and 25\degr~inclination 
and a median mass of 0.20 M$_\sun$ for 60\degr~inclination. 

\begin{figure}
\centering
\vspace{-3cm}
\includegraphics[width=0.7\hsize,angle=0]{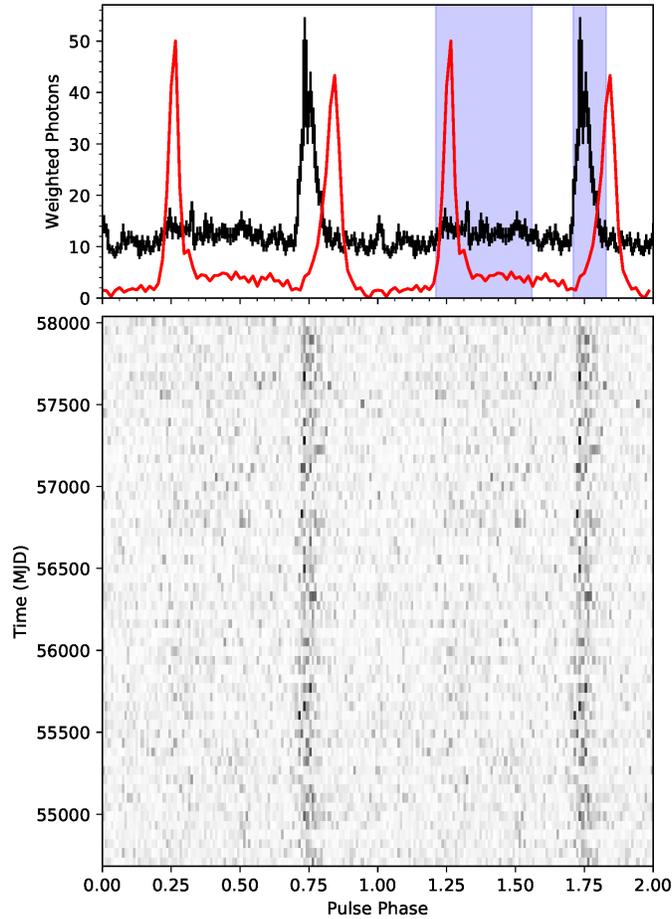}
\caption{Top panel shows a 322 MHz radio profile (red) of PSR J0514$-$4408 plotted with the LAT $\gamma-$ray profile (using $\sim$9.2 years of \textit{Fermi} Large Area Telescope (LAT) Pass 8 data above 100 MeV). Bottom panel shows the LAT phasogram of PSR J0514$-$4408. The uncertainty on the DM in Table \ref{tab:residuals} is 0.006.  At 322 MHz, this corresponds to an error in the DM delay to infinite frequency of 240 $\mu$s (7$\times$10$^{-4}$ of a pulse period)}
\label{fig:aligned_J0514}
\vspace{-2cm}
\end{figure}
Timing residuals for PSRs J0514$-$4408 and J2144$-$5237 contain measurements from simultaneous timing observations 
using the 32 MHz legacy system (black points) and the 200 MHz upgraded GMRT (red points), allowing us to determine the 
timing offset of 1.090011(1) seconds between them. This offset is also validated for two other GHRSS pulsars, 
PSRs J0418$-$4154 and J0702$-$4956, for which the timing solution is reported in P1. The timing offset of the legacy GMRT 
system is well characterized with the other telescopes like Parkes \citep{roy15}, GBT, Lovell, Effelsberg \citep{Dolch14}. 
Knowing the precise timing offset between 32 MHz legacy system and 200 MHz upgraded GMRT will allow combination of the upgraded GMRT data 
with data from other telescopes for high precision timing studies. 

  \begin{figure}[!ht]
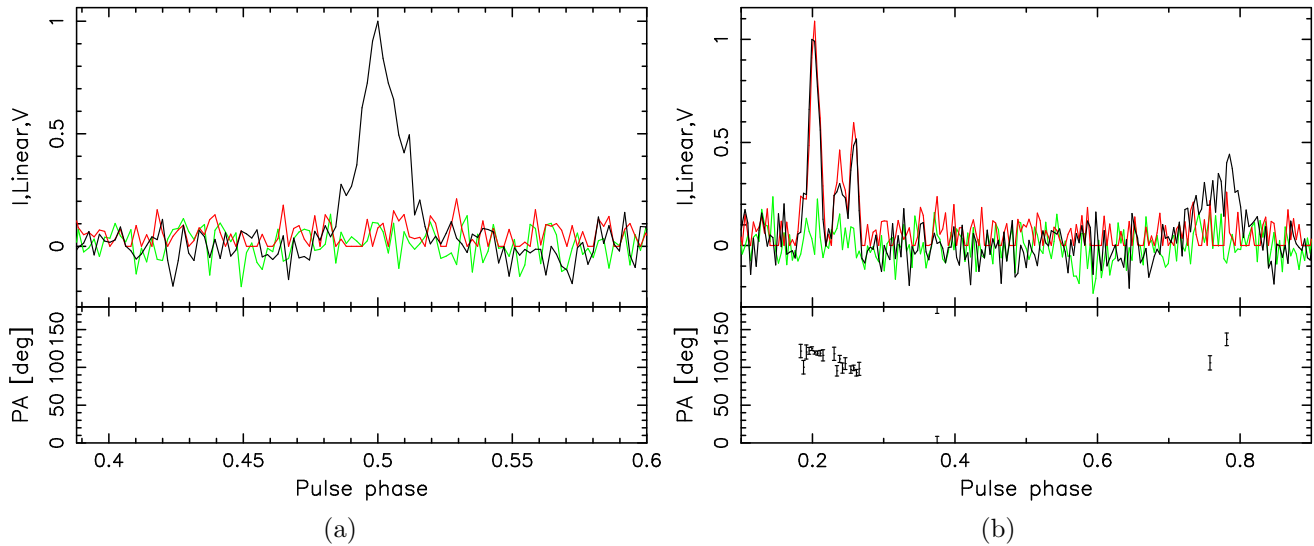

    \subfloat[\label{fig:J0418}]{%
      \includegraphics[width=0.4\hsize,angle=270]{Fig6a.ps}
    }
    \hfill
    \subfloat[\label{fig:J0514}]{%
      \includegraphics[width=0.4\hsize,angle=270]{Fig6b.ps}
    }
    \caption{(a) Polarization pulse profile of for a 3600\,s Parkes telescope observation of PSR~J0418$-$4154 at 1369 MHz center
frequency with 256\,MHz of bandwidth. (b) As (a) for a 13500\,s observation of PSR~J0514$-$4408. In the top panel, the black solid
line represents Stokes $I$, the red line represents the linear polarization, while the green line represents the circular polarization.
The second panel shows the average position angle of the linear polarization (for all the phase bins where the linear polarization
exceeds 2$\sigma$).}
    \label{fig:pol}
  \end{figure}

PSR J1516$-$43 is a mildly recycled pulsar in a binary orbit with an orbital period of $\sim$228 days. 
We determined best-fit barycentric periods at various epochs using 
{\sc psrtime}\footnote{http://www.jb.man.ac.uk/pulsar/observing/progs/psrtime\_commands.html}.
The time variations of these barycentric periods are fitted with a binary model using {\sc fitorbit}, 
to derive a best-fit orbital solution. The periods and model are shown in the left panel of Fig. \ref{fig:1516}. 
The timing model for this pulsar is presented in Table \ref{tab:residuals_1516}, but we are yet to arrive at a 
long-term phase-connected timing solution due to large positional uncertainty ($\pm$ 40\arcmin). The mass 
function of 0.0225 M$_\sun$ corresponds to a companion mass range of 0.42--1.29 M$_\sun$ and a median mass of 0.5 M$_\sun$ for the same inclination angle ranges as above.
\begin{table}
\begin{center}
\caption{$\gamma-$ray spectral fit results for PSR J0514$-$4408\label{tab:latspec}}
\vspace{0.3cm}
\label{tab:latspec}
\begin{tabular}{|l|c|c|c|c|c|c|c|c|c|c|c|c|c|c|c|c}
\hline
Parameter    & Phase-averaged  & Peak 1 & Peak 2 \\\hline
$N_{0}$ (10$^{-11}$ cm$^{-2}$ s$^{-1}$ MeV$^{-1}$) & 1.20$\pm$0.28 & 0.87$\pm$0.18 & 0.36$\pm$0.16\\
$\Gamma$ & 0.77$\pm$0.34 & 0.46$\pm$0.27 & 1.20$\pm$0.50\\
$E_{\rm C}$ (MeV) & 560$\pm$120 & 500$\pm$80 & 590$\pm$230\\
$F$ (10$^{-9}$ cm$^{-2}$ s$^{-1}$) & 7.3$\pm$1.2 & 4.1$\pm$0.4 & 3.1$\pm$0.9\\
$G$ (10$^{-12}$ erg cm$^{-2}$ s$^{-1}$) & 4.8$\pm$0.4 & 3.0$\pm$0.2 & 1.6$\pm$0.3\\
TS & 409 & 883 & 94\\
TS$_{\rm cut}$ & 178 & 153 & 46\\
TS$_{\rm{b\ free}}$ & 1 & 0 & 1 \\\hline
\end{tabular}
\end{center}
Note: Column 1 reports results for the phase-averaged analysis described in Section \ref{latfits}.  Columns 2 and 3 report results for the phase-resolved analyses for each peak in the $\gamma$-ray light curve as described in Section \ref{lattime}. The photon and energy fluxes reported in rows 4 and 5 are integrated from 0.1 to 100 GeV.  All uncertainties are statistical only.
\end{table}

\subsection{Discovery of $\gamma$-ray pulsations from PSR J0514$-$4408}
\label{sec:lat_0514}
\subsubsection{LAT data selection and preparation}\label{latdata}
We selected Pass 8 LAT data spanning the time range from the start of science operations on 2008 August 4 up to 2017 October 12.  We kept events belonging to the SOURCE class, as defined by the \texttt{P8R2\_SOURCE\_V6} instrument response functions, with reconstructed directions within a 15$^{\circ}$ radius of the timing position of PSR J0514$-$4408; energies from 0.1 to 100 GeV; and zenith angles $\leq$ 90$^{\circ}$ to reduce contamination of Earth limb $\gamma$ rays.  We filtered the events to keep only data flagged as good and recorded when the LAT was in nominal science operations mode.  All analyses of \textit{Fermi} LAT data were done using v10r00p05 of the \textit{Fermi} ScienceTools\footnote{Available for download at\\ \url{https://fermi.gsfc.nasa.gov/ssc/data/analysis/software/}.}.

\begin{figure}
\centering
\vspace{-2cm}
\includegraphics[width=0.7\hsize,angle=-90]{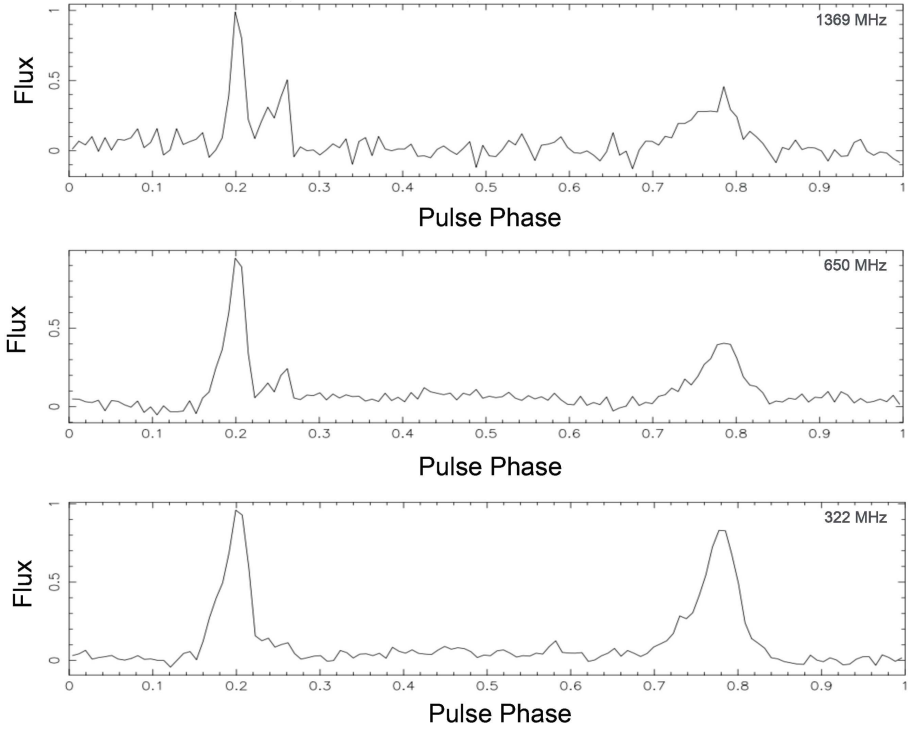}
\caption{The normalized intensity profile of PSR~ J0514$-$4408 at various frequencies. The profiles at 322 and 650 MHz were observed with the GMRT, while the profile at 1369 MHz was observed with the Parkes telescope.}
\label{fig:J0514_freq}
\end{figure}
We constructed a spectral and spatial model of the region by including all sources from the third \textit{Fermi} LAT catalog \citep[3FGL,][] {3FGL} within 25$^{\circ}$ of PSR J0514$-$4408.  For sources within 6$^{\circ}$ of PSR J0514$-$4408, we allowed the spectral parameters to be free only if they were found with an average significance of at least 5$\sigma$.  We also allowed the spectral normalizations of sources within 8$^{\circ}$ of PSR J0514$-$4408 to vary if their 3FGL variability index surpassed the threshold for variability defined by \citet{3FGL}.  The Galactic diffuse emission was modeled using the spectral and spatial template \texttt{gll\_iem\_v06.fits}, while the isotropic diffuse and misidentified cosmic-ray background emissions were modeled jointly with the spectral template\footnote{Both background templates are available for download at\\ \url{https://fermi.gsfc.nasa.gov/ssc/data/access/lat/BackgroundModels.html}.} \texttt{iso\_P8R2\_SOURCE\_V6\_v06.txt} \citep{P8bkgds}.  We moved the 3FGL source associated with PSR J0514$-$4408 to the timing position keeping all the other sources in original 3FGL positions.

The 3FGL catalog was constructed using Pass 7 reprocessed data and covers only the first four years of the mission.  As such, it was necessary to check our initial fits (see Section \ref{latfits}) for new sources in residual counts and test-statistic \citep[TS,][]{3FGL} maps, where the TS is defined as twice the difference in log-likelihood when comparing the fit without the source in the model to the fit with the source. Doing so, we found that it was necessary to free the spectral normalizations of several sources that did not meet our initial criteria (namely, 3FGL J0438.8$-$4519, J0515.3$-$4557, J0533.8$-$3754, J0550.3$-$4521, and J0428.6$-$3756) and that we needed to add two additional sources not in the 3FGL catalog.  One of the new sources was reported already as 2FAV J0451$-$46.8 \citep{2FAV}.  GeV emission from the other new source, associated with PKS~0438$-$43, was first reported in an Astronomer's Telegram \citep{PKS0438ATel}, with a more detailed analysis in a forthcoming paper (Cheung et al.~\emph{in preparation}).

\subsubsection{LAT spectral analysis}\label{latfits}
We performed a binned maximum likelihood spectral analysis on a $20^{\circ}\times20^{\circ}$ region, using the data and region model described in Section \ref{latdata}. The spectrum of PSR J0514$-$4408 was modeled as a power law:
\begin{equation}\label{eq:pl}
\frac{dN}{dE}\ =\ N_{0}\ \Big(\frac{E}{E_{0}}\Big)^{-\Gamma},
\end{equation}
\noindent{}a power law with a simple exponential cutoff:
\begin{equation}\label{eq:co}
\frac{dN}{dE}\ =\ N_{0}\ \Big(\frac{E}{E_{0}}\Big)^{-\Gamma}\ \exp\Big\lbrace-\frac{E}{E_{\rm C}}\Big\rbrace,
\end{equation}
\noindent{}and a power law with a super- or sub-exponential cutoff:
\begin{equation}\label{eq:seco}
\frac{dN}{dE}\ =\ N_{0}\ \Big(\frac{E}{E_{0}}\Big)^{-\Gamma}\ \exp\Big\lbrace-\Big(\frac{E}{E_{\rm C}}\Big)^{b}\Big\rbrace.
\end{equation}

In Equations \ref{eq:pl}-\ref{eq:seco}, $N_{0}$ is a normalization parameter with units of MeV$^{-1}$ cm$^{-2}$ s$^{-1}$; $E_{0}$ is a scale parameter, chosen to be 534 MeV, the \texttt{PIVOT\_ENERGY} of the corresponding 3FGL source; and $\Gamma$ is the photon index.  In Equations \ref{eq:co} and \ref{eq:seco} $E_{\rm C}$ is the cutoff energy, and the $b$ parameter in Equation \ref{eq:seco} is the exponential index, which controls the strength of the spectral cutoff. Note that fixing $b=1$ returns Equation \ref{eq:co}.  The spectra of most $\gamma$-ray pulsars are well described by Equation \ref{eq:co}, but the spectra of the brightest pulsars detected with the LAT are better fit with Equation \ref{eq:seco} with $b<1$, suggesting a sub-exponential cutoff \citep{2PC}.

\begin{table}
\begin{center}
\caption{$\gamma-$ray pulse profile details for PSR J0514$-$4408}
\vspace{0.3cm}
\label{tab:latprof}
\begin{tabular}{|l|c|c|c|c|c|c|c|c|c|c|c|c|c|c|c|c}
\hline
Parameter & Value \\\hline
$\phi_{1}$ & 0.735$\pm$0.003\\
$w_{1,l}$ & 0.014$\pm$0.002 \\
$w_{1,t}$ & 0.034$\pm$0.0030\\
$\phi_{2}$ & 0.393$\pm$0.021\\
$w_{2}$ & 0.137$\pm$0.017\\\hline
\end{tabular}
\end{center}
Note: The peak widths ($w_{i}$) are given as half-width at half max values.  For the two-sided Gaussian, $w_{1,l}$ refers to the leading side of the peak, earlier phases, while $w_{1,t}$ refers to the trailing side of the peak, later phases.
\end{table}

To test which model was preferred, we followed \citet{2PC} and computed TS$_{\rm cut}$ to compare fits using Equations \ref{eq:pl} and \ref{eq:co} and TS$_{\rm{b\ free}}$ 
to compare fits using Equations \ref{eq:co} and \ref{eq:seco}. The likelihood showed a significant preference for the simple exponential cutoff over the power law, and no strong preference for the fit with $b$ as a free parameter.  The phase-averaged best-fit spectral values, point source TS, TS$_{cut}$, TS$_{b\ free}$, and integrated photon ($F$) and energy ($G$) fluxes are given in column 1 of Table \ref{tab:latspec}.  The phase-averaged $\gamma$-ray spectrum is shown in Fig. \ref{fig:latspectrum}.
The phase-averaged values of $\Gamma$ and $E_{\rm C}$ reported in Table \ref{tab:latspec} are consistent with those of other $\gamma$-ray pulsars in \citet{2PC} with similar characteristics. PSR J0514$-$4408 occupies a region of low spin-down power primarily populated by millisecond pulsars, but our best-fit $\Gamma$ value is similar to those sources, further supporting that the same emission mechanism is operating in young and recycled $\gamma$-ray pulsars. The magnetic field strength at the light cylinder of PSR J0514$-$4408, $B_{\rm LC}\sim230 G$, is relatively weak compared to other known $\gamma$-ray pulsars, and our value of $E_{\rm C}$ agrees well with the trend of cutoff energy with $B_{\rm LC}$ noted by \citet{2PC}.
 
\subsubsection{LAT timing analysis}\label{lattime}
We folded the $\gamma$-ray data using the radio timing solution described in Section \ref{sec:timing}. Then, using the best-fit model from 
Section \ref{latfits}, we computed spectral weights for each event within 3$^{\circ}$ of PSR J0514$-$4408, reflecting the probability that 
a given event should be associated with the pulsar. These spectral weights have been shown to enhance the detectability of pulsations in 
LAT data \citep{Kerr11}. Using the phase-folded and spectrally weighted events with a weighted version of the H test \citep{deJager89,deJager10,Kerr11}, 
we detected significant pulsations. However, analysis of the pulse phase vs.~time suggested that the timing solution did not accurately 
describe the pulsar rotation at epochs preceding the radio discovery. Fig. \ref{fig:aligned_J0514} presents the phasogram for PSR J0514$-$4408. 
An aligned 322 MHz profile and the $\gamma$-ray LAT profile after refining the timing solution as described below, is presented in the top panel, 
corresponding to a detection significance of 32$\sigma$.

We fit an analytic template to the pulse profile comprising a two-sided Gaussian for the main peak and a single-sided Gaussian for the broad, 
second peak. We then extracted 50 TOAs by cross-correlating with the unbinned data \citep{Ray11}, and we then produced a joint timing solution 
by fitting the model parameters to the radio and \textit{Fermi} TOAs with \textsc{Tempo2}. We subsequently iterated the process to produce a 
final analytic template and timing model\footnote{The final timing solution will be available at \\ \url{https://fermi.gsfc.nasa.gov/ssc/data/access/lat/ephems/}.}.  
With the final timing solution, we find that a two-peak pulse profile model is strongly preferred, with a log likelihood difference 
between a one- and two-peaked model of 21.39 for three extra degrees of freedom.  The best-fit parameters of the $\gamma$-ray pulse profile 
are given in Table \ref{tab:latprof}. 

In order to estimate the inclination angle of its magnetic axis and that of our line of sight with respect to the spin 
axis of PSR J0514$-$4408, we used the geometric models and fitting methods of \citet{Johnson2014} 
and \citet{Wu2018}.  Our initial attempts, assuming a simulated spin period of 100 ms, period derivative 
of 1$\times10^{-15}$ s s$^{-1}$, and radio frequency of 300 MHz; were unsuccessful at simultaneously matching the 
observed $\gamma$-ray and radio pulse profiles. However, as discussed in Section \ref{subsec:j0514pol}, there is reason 
to think that the radio emission originates at a relatively high altitude in the magnetosphere ($\gtrsim1130$ km).  The 
simulated radio pulse profiles we use follow the model of \citet{Story2007}, which \citep[using the radius to frequency mapping of][]{KG2003} 
place the 300 MHz emission at an altitude of 270 km (assuming a neutron star radius of 10 km). We generated new simulated 
radio pulse profiles, with emission at an altitude of 1130 km, and performed new fits. These new fits matched the $\gamma$-ray 
profiles well for moderate inclination and viewing angles, but the fits to the radio profile were still not satisfactory and 
suggested a higher altitude might still be needed.






In order to investigate the spectral behavior in each peak, we employed a modified Bayesian blocks \citep{Scargle89,Jackson05,Scargle13} analysis to define the relevant phase intervals for the $\gamma$-ray pulse profile, with 100 phase bins per rotation.  Our method used the weighted counts profile, similar to what was done by \citet{CaliandroJ0632} but using a weighted average when deciding whether or not to split a block.  We reproduce the $\gamma$-ray pulse profile over the phase range $-$1 to 2 and use an f-test, with a $\chi^{2}$ statistic, to test if the data are best described by one block or two, requiring that the probability of incorrectly splitting one block into two be $\leq$ 0.05.  If we split the block into two, we then test the leftmost (earlier phase) block to see if it should be split or not, stopping when the f-test suggests one block is sufficient or the block reaches the minimum size we impose of 5 phase bins.  This process moves to later phases and continues until there are no more blocks to test.

Using this method, we defined three phase intervals of interest.  We denoted the highest $\gamma$-ray peak as peak 1, defined to be phases $\phi\ \in\ [0.71,0.83]$.  A broader, but lower amplitude peak 2 is defined as $\phi\ \in\ [0.21,0.58]$. These two peaks are indicated by the blue shaded 
region in Figure \ref{fig:aligned_J0514}. The final phase interval, the off-peak, was defined as $\phi\ \in [0.0,0.21)\cup(0.58,0.71)\cup(0.83,1.0)$.  We performed binned maximum likelihood fits in each phase interval, accounting for the difference in exposure.  We started with the best-fit model from the phase-averaged analysis described in Section \ref{latfits}, but only let the normalizations of other sources within 3$^{\circ}$ be free to vary. The normalization of the isotropic diffuse component was allowed to vary, but the spectral parameters of the Galactic diffuse component were held fixed.  The results for both peak 1 and peak 2 are given in columns 3 and 4 of Table \ref{tab:latspec}, with the likelihood strongly favoring the power-law with a simple exponential cutoff shape in both cases.

The best-fit photon index of peak 1 is harder than peak 2, but the difference is not significant.  Aside from the flux difference, the spectra in the peaks are compatible.  There was no significant detection in the off-peak phase interval, TS = 5 assuming a power-law shape.  We calculate a 95\% confidence-level upper limit on the integrated photon flux, 0.1 to 100 GeV, of $5.3\times10^{-9}$ cm$^{-2}$ s$^{-1}$ on any non-pulsed $\gamma$-ray emission, correcting the exposure to a phase-averaged value.

From the best-fit spectrum for the phase-averaged $\gamma$-ray data, we can 
derive a $\gamma$-ray luminosity of $L_{\gamma}\ =\ 4\pi\ f_{\Omega}\ G\ d^{2}\ =\ (9.61\pm0.85)\times10^{32}$ erg s$^{-1}$,
using a beaming factor of $f_{\Omega}\ \sim\ 1$ \citep{2PC}, the DM-derived distance ($d$), and the energy flux ($G$).
Assuming a neutron star moment of inertia of $1\times10^{45}$ g cm$^{2}$, the timing parameters in Table \ref{tab:residuals} yield a spin-down power
for PSR J0514$-$4408 of $\dot{E}\ =\ 2.45\times10^{33}$ erg s$^{-1}$.
This implies an efficiency of converting rotational energy into $\gamma$-rays of $\eta_{\gamma}\ =\ L_{\gamma}/\dot{E}\ =\ 0.39\pm0.03$,
which agrees well with other known $\gamma$-ray pulsars with similar $\dot{E}$ \citep{2PC}.

\subsection{Polarization and multi-frequency study}
\label{sec:polarization}
\subsubsection{PSR J0418$-$4154}
The polarimetric pulse profile of PSR J0418$-$4154 at a center frequency of 1369 MHz and a bandwidth of 256 MHz obtained from 
an observation of 3600 s duration with the Parkes telescope is displayed in Fig. \ref{fig:J0418}. This pulsar appears to be 
unpolarized at our observing frequency. Fitting for Rotation Measure (RM) was attempted but no significant value was measured.\\

\subsubsection{PSR J0514$-$4408}\label{subsec:j0514pol}
The polarization pulse profile of PSR J0514$-$4408 at 1369 MHz is displayed in Fig. \ref{fig:J0514}. The main pulse (MP), 
which peaks at pulse phase 0.2, appears to be completely linearly polarized, with no significant circular polarization, while 
the interpulse (IP) is unpolarized. The PA swing in the MP is relatively shallow and smooth, with no orthogonal polarization 
mode (OPM) jumps. Given the lack of detail in the shape of the PA swing and the fact that we only detect enough significant 
PA points for the MP, fitting the Rotating Vector Model \citep[][RVM]{rc69} to the swing was not constraining.

We can infer more about the geometry of this pulsar by looking at the frequency evolution of the total intensity profile, 
displayed in Fig. \ref{fig:J0514_freq}. The profiles at 322, 650 and 1369 MHz were aligned using the leftmost component 
of the MP, so that this feature was always at phase 0.2. The change in profile shape at phase 0.18, seen at all 
frequencies, indicates that we correctly identified and aligned the same components in the plots.

Although we observe a MP and IP, it does not automatically follow that the pulsar is an orthogonal rotator, and that what we 
observe is radiation coming from both magnetic poles. The separation between the MP and IP is more than 0.5 phase. Thus, it 
could be the two sides of a wide beam centered on a magnetic axis at a low inclination angle of the magnetic axis relative to 
the rotation axis. This could be taken as evidence in favor of a wide profile and a low magnetic 
inclination angle. The strong indicator of it being an orthogonal rotator, however, is that the separation between the two peaks 
remains constant as a function of frequency. In general the emission height is thought to be a function of frequency, which 
should change the separation of the components observed from the same pole \citep{lorimer04}. For this pulsar, we see significant 
profile evolution with frequency but the separation between the MP and the IP remains unaffected. Another 
indication of it being an orthogonal rotator is the lack of significant bridge emission in between the MP and IP at any 
frequency. We do not see any sign of bridge emission being present in 322, 650 and 1369 MHz, although our flux limit is not particularly 
constraining. 

\begin{figure}
\centering
\includegraphics[width=0.4\hsize,angle=270]{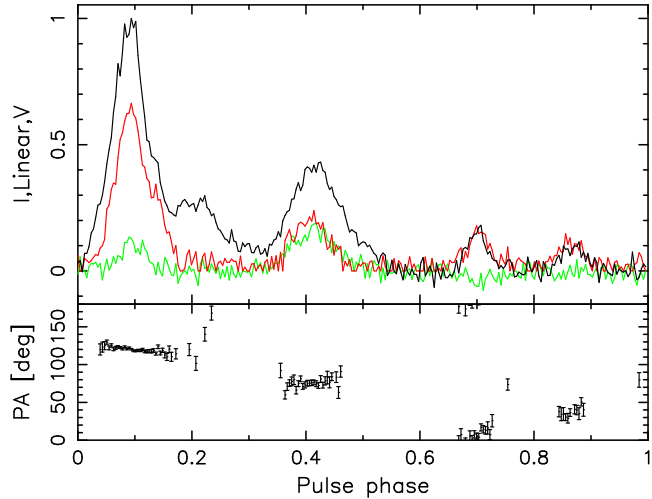}
\caption{Polarization pulse profile for a 12250\,s observation of PSR~ J2144$-$5237, as Figure \ref{fig:J0418}.}
\label{fig:J2144}
\end{figure}

\begin{figure}
\centering
\vspace{-1cm}
\includegraphics[width=0.7\hsize,angle=-90]{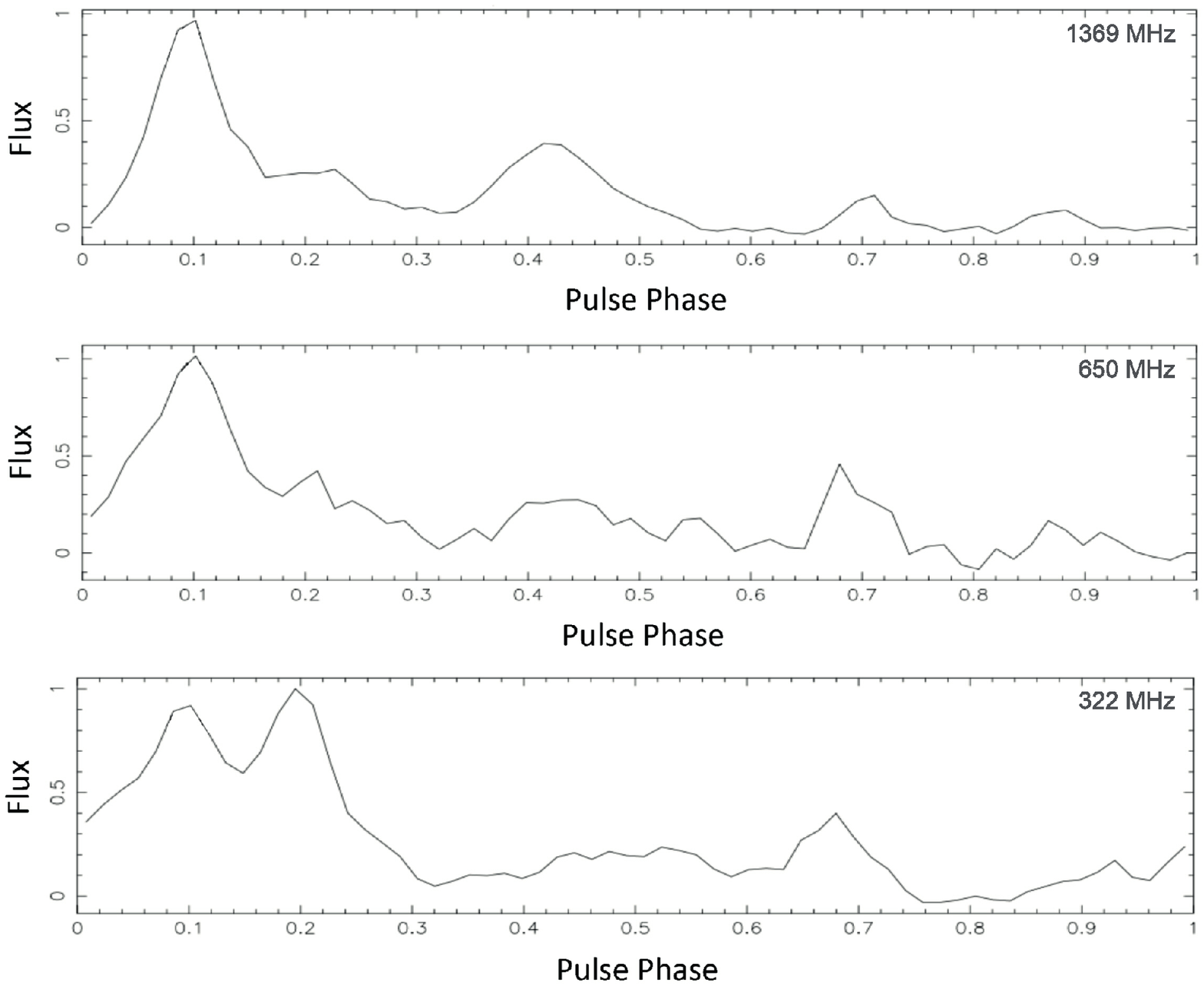}
\caption{The normalized intensity profile of PSR~ J2144$-$5237 at various frequencies. The profiles at 322 and 650 MHz were observed with the GMRT, while the profile at 1369 MHz was observed with the Parkes telescope.}
\label{fig:J2144_freq}
\end{figure}

The MP of this pulsar looks like a single peaked component at 322 MHz with weak structure on the trailing edge of the pulse. 
From  Fig. \ref{fig:J0514_freq}, we can see that with increasing frequency the structure evolves into two peaks. This indicates 
that the fiducial plane crossing (the plane containing both the rotation and the magnetic axis) might be at a later pulse phase 
than initially inferred from the 322 MHz profile. Furthermore, by looking at the IP we see that the profile is asymmetrical at 
all observed frequencies, as the leading edge has a slower rise compared to the trailing edge. This might indicate that the 
fiducial plane crossing is at an earlier pulse phase with respect to where the IP peaks. This would explain the deviation 
from 0.5 phase difference between the MP and IP, as should be the case if we see opposite poles of the neutron star.

That PSR J0514$-$4408 may be an orthogonal rotator is consistent with the classical 
$\gamma$-ray models (e.g. \citealt{wrw+09}) predicting that a large fraction of orthogonal rotators can be detected.
Nevertheless, \citet{rwj15a} obtained a distribution of inclination angles for 28 $\gamma$-ray loud pulsars, and 
observed an unexpected skewness towards low values. Despite this skew, a number of pulsars with 
higher $\alpha$ (angle between the rotation and magnetic axis) are present in their sample.
For example, PSRs J0908$-$4913 and J1057$-$5226 from their sample have interpulses and higher values of $\alpha$.

Since the PA swing of the MP is relatively flat, we can measure the steepest gradient by fitting a straight line, and 
we obtain $(-1.1\pm0.1)$. If PSR J0514$-$4408 is an orthogonal rotator, as discussed before, we consider the closest 
approach of the line-of-sight to the magnetic axis to occur somewhere within the MP, and the steepest region of 
the PA gradient (inflection point) to occur in this region as well, as predicted by the RVM. Our value for the gradient 
is actually a lower limit, as the steepest gradient value could occur beyond the MP. Such a delay could arise due to 
relativistic effects \citep{bcw91}. The gradient of the steepest part of the RVM can be written in terms of the magnetic 
inclination angle, $\alpha$, and the angle between the magnetic axis and the line-of-sight at closest approach, 
$\beta$, as $\sin(\alpha)/\sin(\beta)$ \citep{kom70}. For an orthogonal rotator, $\sin(\alpha)$ is close to 1, 
and we can infer an upper limit on $\beta$ as $\sim-65^{\circ}$, since $\beta$ is inversely proportional to the steepest gradient.

A large value of $\beta$ would imply that both emission cones are very wide ($\sim130^{\circ}$). If the beams are fully illuminated, 
it is very hard to explain how the line-of-sight could pass the two cones of emission in such a way as to create both a 
narrow MP and IP.  From Fig. \ref{fig:J0514_freq}, we can estimate the fractional pulse width (full width half maximum) at 1369 MHz 
of the leading component of MP as $W_{\text{MP}}\sim$ 0.07  and of IP as $W_{\text{IP}}\sim$0.14. Hence, as narrower beams and a 
lower value of $\beta$ are needed to explain the observed pulse widths, the 
value of the steepest gradient should be higher than what we observe. 
As stated before, this could be the case if the steepest part of the PA swing occurs at a later pulse phase compared to 
the fiducial plane crossing, somewhere outside the observed MP, at least at 0.05 rotational phase delay. This delay, 
$\Delta\phi$, measured in radians, is predicted to be

\begin{equation}
\Delta\phi = \frac{8 \pi h_{\text{em}}}{Pc},
\label{eq:delta}
\end{equation}

\noindent where $h_{\text{em}}$ is the emission height measured with respect to the centre of the star, $P$ is the period of the pulsar and $c$ represents the speed of light \citep{bcw91}. Using the predicted lower limit on the rotational phase delay in Equation \ref{eq:delta} gives a lower limit on the emission height as $h_{\text{em}} \sim 1130$ km, which is relatively high compared to other $\gamma$-loud pulsars \citep{rwj15a}. From the sample of pulsars presented by  \citet{rwj15a}, only PSR J0659+1414 had a higher emission height compared to our lower limit, indicating that the actual value of $h_{\text{em}}$ is unlikely to be much larger than the limit.

Considering that the emission comes from a cone, which is defined by the last open magnetic field lines, one can define the angular radius of this region, $\theta_{\text{em}}$, as

\begin{equation}
\theta_{\text{em}} = \arcsin{ \left(\sqrt{\frac{2 \pi h_{\text{em}}}{Pc}} \right)}.
\label{eq:polar}
\end{equation}
We can relate $\theta_{\text{em}}$ to the half-opening angle of the radio beam by

\begin{equation}
\rho = \theta_{\text{em}} + \arctan{\left( \frac{\tan(\theta_{\text{em}})}{2} \right)},
\label{eq:beam}
\end{equation}

\noindent (e.g. \citealt{gg01}). Using our predicted lower limit on $h_{\text{em}}$ in Equation \ref{eq:polar}, and substituting the resulting value of $\theta_{\text{em}}$ in Equation \ref{eq:beam}, we infer a lower limit on the half-opening angle of the beam to be $\rho\sim25^{\circ}$. This implies that the two emission beams could be at least $50^{\circ}$ wide, which could explain the observed narrow MP and IP. Given that the two radio beams of this pulsar are still relatively wide, $\alpha$ does not have to be exactly $90^{\circ}$ in order to explain the observed pulse widths.
We measured a value of $(17.3 \pm 5.9)$ rad m$^{-2}$ for the RM for this pulsar using a method based on the RM synthesis technique, described in more detail by \citet{ijw18}.\\

\subsubsection{PSR J2144$-$5237}
The polarization pulse profile of PSR J2144$-$5237 at 1369 MHz is displayed in Fig. \ref{fig:J2144}. The profile appears to span the whole pulse 
phase consisting of five clear components. Such a wide multi-component profile is seen for some MSPs (e.g. \citealt{dhm+15}).
The degree of linear polarization is relatively high for all components, except for the second component, at rotational phase 0.2,  which appears to be unpolarized. The PA swing has a complex behaviour with different gradients and slope signs for different components, making it  impossible to fit with the RVM. There are no observed OPM jumps visible in the shape of the PA curve. In general, the PA curves of MSPs do not fit the RVM model very well.
However, \citet{dhm+15} managed to fit the RVM model for some pulsars in their sample using their higher frequency observations. Deviations from the model occurred when using lower frequencies, and the authors suggested that a reason for this could be that lower frequencies are generated further away from the neutron star surface. Because the magnetospheres of MSPs are so compact, these frequencies are probably generated close to the light-cylinder radius where the magnetic field deviates from a dipolar field.

The total intensity profile of PSR J2144$-$5237 at three different frequencies (322, 650 and 1369 MHz) is displayed in Fig. \ref{fig:J2144_freq}. The profiles were aligned based on the first (at pulse phase 0.1) and fourth (at pulse phase 0.7) components, which could be clearly identified at all frequencies and their separation did not change as a function of frequency. A slower evolution of the pulse profile with frequency, especially component separation, is observed in several MSPs (compared to normal pulsars), also likely due to their very compact magnetospheres \citep{kll+99}. In Fig. \ref{fig:J2144_freq}, the most striking evolution with frequency can be seen for the second component, situated at pulse phase 0.2.
The RM value measured for this pulsar is $(25.1 \pm 1.9)$ rad m$^{-2}$.

\section{Summary}
\label{sec:discussion}
In this paper we present the discovery of three pulsars with the GHRSS phase-1 survey, a timing study of three of the 
other newly discovered pulsars, the discovery of $\gamma$-ray pulsations from one of the GHRSS pulsars, and polarization 
results for three GHRSS pulsars. 

We report the discovery of PSRs J1239$-$48, J1516$-$43 and J1726$-$52. Our estimates of the flux densities for the newly 
discovered pulsars suggests that we are achieving our theoretical sensitivity limit of $\sim$0.5 mJy. The discovery of 13 
pulsars in GHRSS phase-1 from 1800 square degree sky coverage indicates a discovery rate of 0.007 pulsars per square degree, 
which is one of the highest among the off-Galactic plane surveys (e.g. \cite{stovall14}). 
 
We present timing models for PSRs J2144$-$5237, J0514$-$4408 and J1516$-$43 which were discovered with the GHRSS survey. 
PSR J2144$-$5237~is a millisecond pulsar with period of 5.04~ms~in a 10~day orbit around a $\leq$0.18~M$_{\sun}$ companion. 
PSR J0514$-$4408 discovered with the GMRT has a period of 320.27~ms and is associated with a \textit{Fermi} LAT source 
emitting $\gamma$-ray pulses. The relative phase alignment between the $\gamma$-ray and radio light curve is intriguing and 
should provide meaningful insight into emission models for both wavelengths. A more detailed investigation is deferred to 
a future paper. The spin period of PSR J1516$-$43 is in between the bulk of the normal pulsars and millisecond pulsars and is in a 
wide binary (orbital period of $\sim$228~days). The range of possible companion masses of 0.42$-$1.29~M$_{\sun}$ implies 
that the companion could either be a white-dwarf (for low inclination angles) or a low-mass neutron star (for higher inclination angles). 
Considering the typical mass range of white-dwarfs and neutron stars and the most likely inclination angles, we calculate a $\sim$75 \% 
probability that the companion will be a white dwarf and $\sim$25\% probability that the companion will be neutron star. If it were
a low-mass neutron star companion with a very low orbital eccentricity then the second-born neutron star would have to have received 
a very small velocity kick at birth \citep{vdh07}.
Comparing the rotational periodicities of the fully recycled MSPs and the normal pulsars, as well as the companion mass, suggests 
that PSR J1516$-$43 is a mildly recycled pulsar. It is thought that in the recycling process MSPs with massive CO/ONeMg white dwarfs 
are often mildly recycled with 10 $< P <$ 100 ms and 10$^{-20} < \dot{P} <˙10^{-18}$ \citep{tauris12}.

Folding $\sim$9.2 years of \textit{Fermi} Large Area Telescope Pass 8 data above 100 MeV using the radio timing ephemeris derived from the GMRT 
observations, we found a 32$\sigma$ detection of $\gamma$-ray pulsations from PSR J0514$-$4408. 
Our best-fit spectral properties and derived $\gamma$-ray efficiency agree well with those of other $\gamma$-ray 
pulsars with similar characteristics. A phase-resolved analysis of the LAT data reveals no evidence for off-peak emission, 
such as what might be expected from a pulsar wind nebula.

We study the profile evolution for PSR J0514$-$4408 between 322, 650 and 1369 MHz. Similar 
MP and IP strengths are observed at 322 MHz, whereas at 650 and 1369 MHz the relative strength of the IP decreases. We report high 
linear polarization for the MP and unpolarised IP at 1369 MHz. We infer that PSR J0514$-$4408 is possibly an orthogonal rotator 
and discuss the consequences.   

MSP J2144$-$5237 has a wide multi-component pulse profile. We report significant 
linear polarization for most of the profile components. A slower evolution of the pulse profile with frequency is 
observed for MSP J2144$-$5237 than is commonly seen in some MSPs (Kramer et al. 1999). However, one 
of the profile components with small linear polarization seems to exhibit significant profile evolution 
between 322 to 1390 MHz. 

\acknowledgments

The \textit{Fermi} LAT Collaboration acknowledges generous ongoing
support from a number of agencies and institutes that have supported
both the development and the operation of the LAT as well as
scientific data analysis.  These include the National Aeronautics and
Space Administration and the Department of Energy in the United
States, the Commissariat \`a l'Energie Atomique and the Centre
National de la Recherche Scientifique / Institut National de Physique
Nucl\'eaire et de Physique des Particules in France, the Agenzia
Spaziale Italiana and the Istituto Nazionale di Fisica Nucleare in
Italy, the Ministry of Education, Culture, Sports, Science and
Technology (MEXT), High Energy Accelerator Research Organization (KEK)
and Japan Aerospace Exploration Agency (JAXA) in Japan, and the
K.~A.~Wallenberg Foundation, the Swedish Research Council and the
Swedish National Space Board in Sweden.
 Additional support for science analysis during the operations phase
is gratefully acknowledged from the Istituto Nazionale di Astrofisica
in Italy and the Centre National d'\'Etudes Spatiales in France. This
work performed in part under DOE Contract DE-AC02-76SF00515.

B. Bhattacharyya acknowledges support of Marie Curie grant PIIF-GA-2013-626533 of European Union.  
Pulsar research at Jodrell Bank Centre for Astrophysics and Jodrell Bank Observatory is supported by 
a consolidated grant from the UK Science and Technology Facilities Council (STFC). 
B. W. Stappers and S. Cooper acknowledge funding from the European Research Council (ERC) under the European Union's 
Horizon 2020 research and innovation programme (grant agreement No. 694745). M. A. McLaughlin is a member of the 
NANOGrav Physics Frontiers Center, supported by NSF award number 1430284. She is also supported by NSF award number 1458952.
Work at NRL is supported by NASA. We thank Lucas Guillemot for reviewing our paper as LAT internal reviewer and providing 
very useful suggestions. We also thank Dave Thomson and Gulli Johannesson for their comments on the draft. 
The Parkes radio telescope is part of the Australia Telescope National Facility which is funded by the 
Australian Government for operation as a National Facility managed by CSIRO. 
We thank VVS and S. Gole for maintaining the compute cluster at the NCRA. 
We thank the staff of the GMRT who have made these observations possible. We acknowledge support of GMRT operators. The GMRT 
is run by the National Centre for Radio Astrophysics of the Tata Institute of Fundamental Research.


\end{document}